# VLA Observations of Ultraluminous IRAS Galaxies: Active Nuclei or Starbursts?

Thomas Crawford[1,2], Jonathan Marr[3] and Bruce Partridge[2]
Haverford College, Haverford, Pennsylvania 19041
and
Michael A. Strauss[2,4]
Institute for Advanced Study, Princeton, New Jersey 08540

ABSTRACT

We employed the Very Large Array (VLA) of the National Radio Astronomy Observatory[5] in C configuration to map 39 ultraluminous *IRAS* galaxies at $\lambda = 6$ cm and 20 cm, at resolutions of $\approx 4''$ and $\approx 15''$, respectively. All sources were detected at both frequencies. The sources were selected from the flux-limited redshift survey of Strauss *et al.* (1990; 1992) and comprise nearly half of those sources with far-infrared luminosity $\geq 4.3 \times 10^{11}$ $L_\odot$ (for $H_0 = 75\,\mathrm{km\,s^{-1}\,Mpc^{-1}}$) in the Northern sky. Twenty-four of these sources were later remapped with the VLA in A configuration at $\lambda = 6$ cm with a resolution of $\approx 0\farcs5$.

The fluxes of our sources range from 1.2 to 187 mJy at 6 cm, with the majority of the sources fainter than 20 mJy, and from 4 to 150 mJy at 20 cm, with the majority of the sources fainter than 50 mJy. Most of the sources have radio spectral indices indicative of synchrotron emission ($\alpha \approx -0.65$). There is one source, however, that shows an inverted spectrum with $\alpha = +2.1$; observations at higher frequencies show that the spectrum peaks between 5 and 8 GHz, as high as any of the "gigahertz peaked spectrum" sources studied by O'Dea *et al.* (1990; 1991). We discuss the implications of this source for observations of fluctuations in the cosmic microwave background.

Eighteen of the sources are resolved at $4''$ resolution. The extended sources fall into several categories: two show multiple unresolved components, another four are doubles with at least one resolved component, 14 show extended emission which could arise from a disk, and two show arc-second long jets. We find that our data fit the tight correlation found by Helou *et al.* (1985) between far-infrared and microwave luminosity; this correlation extends to the highest infrared luminosities. We also find that the correlation is weaker if only the extended or the nuclear components of the radio luminosity are used. We therefore argue that the far-infrared emission in the majority of these higher luminosity galaxies is due to the same mechanism as the lower luminosity FIR sources, which is believed to be star formation, rather than non-thermal activity in the nucleus. Moreover, the star formation is not confined to the extended disk in these sources, but is important in the nucleus as well.

[1] Current Address: Department of Physics, Campus Box 390, University of Colorado, Boulder, Colorado, 80309

[2] Guest Observer, The Very Large Array

[3] Current Address: Physics Department, Union College, Schenectady, NY 12308

[4] Current Address: Department of Astrophysical Sciences, Princeton University, Princeton, NJ 08544

[5] The National Radio Astronomy Observatory is operated by Associated Universities, Inc., under cooperative agreement with the National Science Foundation.

## 1. Radio Emission from Star-forming Galaxies

One of the most dramatic early discoveries by the *Infrared Astronomical Satellite*[6] (*IRAS*) was the existence of numerous galaxies with extremely high far-infrared (FIR) to optical emission ratios; these objects have FIR luminosities up to $10^{12}$ $L_\odot$ (Soifer, Houck, & Neugebauer 1987a). The physical origin of this tremendous source of energy remains unclear, with various authors suggesting starbursts (e.g., Condon *et al.* 1991c), embedded active galactic nuclei (e.g., Sanders *et al.* 1988, 1989), and kinetic energy from galaxy mergers (Harwit *et al.* 1987). At lower FIR luminosities, it is well established that the majority of FIR emission is due to star formation and the heating of interstellar dust (FIR cirrus) by the ambient radiation field. Evidence for extensive star formation in these galaxies is reviewed in Soifer *et al.* (1987a) and is based on observations of optical emission lines, especially H$\alpha$, a tracer of star formation (e.g., Kennicutt 1983), near-infrared observations of CO absorption due to young supergiants (cf., Ridgway, Wynn-Williams, & Becklin 1994; Smith *et al.* 1995), direct imaging of supernova remnants in nearby infrared galaxies such as M82 (Huang *et al.* 1994), and model fits involving star-formation to the spectral energy distributions of these galaxies (e.g., Rowan-Robinson & Crawford 1989). However, at the most extreme luminosities observed, an appreciable fraction of the sources show evidence for AGN (active galactic nuclei) in their optical spectra (cf., Sanders *et al.* 1988; Armus *et al.* 1995). Ultraluminous *IRAS* galaxies are by their nature dusty systems, so their central regions are difficult to study optically. It thus remains unclear from optical observations whether these AGN are ubiquitous, or indeed, whether the AGN that are seen are capable of providing the prodigious energy output seen in the FIR.

Observations at radio wavelengths, however, penetrate the optically obscuring dust and can look directly into the central regions of these galaxies. We have undertaken a study of a sample of ultraluminous *IRAS* galaxies at 6 cm (C band) and 20 cm (L band) with the Very Large Array (VLA). Radio observations can give us several insights into the physical processes of these galaxies:

a. There is a remarkable observed correlation between the FIR and microwave luminosities of galaxies (cf., Dickey & Salpeter 1984; de Jong *et al.* 1985; Sanders & Mirabel 1985; Helou, Soifer, & Rowan-Robinson 1985; Wunderlich, Weilebinski, & Klein 1987; Unger *et al.* 1989; Condon & Broderick 1991; cf., the review in Condon 1992), suggesting a direct relation between the sources of the radio and FIR emission.
b. The high angular resolution provided by the VLA allows us to use source morphology to discriminate between central AGN emission (and possible associated features, such as jets) and more diffuse emission from surrounding star formation regions.

---

[6] The Infrared Astronomical Satellite was developed and operated by the U.S. National Aeronautics and Space Administration (NASA), the Netherlands Agency for Aerospace Programs (NIVR), and the U.K. Science and Engineering Research Council (SERC).

c. Measurements of the radio spectral indices of these galaxies can distinguish between various possible emission mechanisms.

### *1.1 FIR-Radio Correlation*

The linear relation between FIR and microwave emission can be expressed in terms of flux densities as

$$q \equiv \log(S_{\rm FIR}/S_{20\,{\rm cm}}), \tag{1}$$

where

$$S_{\rm FIR} = 0.336\,(2.58 S_{60\,\mu{\rm m}} + S_{100\,\mu{\rm m}}) \tag{2}$$

is the effective flux density at $80\,\mu$m (Lonsdale *et al.* 1985), and $q$ is constant at $\approx 2.34$ (e.g., Condon, Andersen, & Helou 1991a). This relation is observed to hold over four decades in FIR luminosities, with a scatter of only $\approx 0.2$ dex (although there have been claims of non-linearities in this relationship; c.f. Fitt, Alexander & Cox 1988; Devereaux & Eales 1989; Chi & Wolfendale 1990; Condon *et al.* 1991a). This correlation is very surprising, as this range in luminosity extends from quiescent spirals through active star-forming spirals, including extreme starburst galaxies, and on to active galactic nuclei and quasars (cf. Sopp & Alexander 1991). Of course, this relation holds only for galaxies with significant FIR emission, as there does exist a population of galaxies with strong radio emission and weak FIR emission, *i.e.*, the classic radio galaxies (e.g., Golembek, Miley, & Neugebauer 1988; Dey, Shields, & van Breugel 1990). Helou & Bicay (1993) propose a model for the infrared-radio correlation in which the starburst activity that gives rise to FIR emission results in supernova explosions, whose remnants emit in the radio via synchrotron radiation. Their model, however, is fit to the radio and infrared luminosity profiles of typical galaxies and does not successfully explain the extension of the FIR-radio correlation to the most luminous galaxies. Other models explaining the correlation in terms of star formation include those of Fitt, Alexander, and Cox (1988), Völk (1989), Chi & Wolfendale (1990), and Colina & Perez-Olea (1992).

On the other hand, Helou *et al.* (1985) find that the correlation has a larger dispersion at higher luminosities and therefore may break down for the highest luminosity *IRAS* galaxies. Harwit *et al.* (1987) argue that the ultraluminous *IRAS* galaxies are fundamentally different in nature from the lower-luminosity systems, as indicated by a break in the FIR galaxy luminosity function at $10^{11} L_\odot$ (Lawrence *et al.* 1986). Citing studies that indicate that the extremely bright *IRAS* galaxies seem to be associated with interacting galaxies, they propose a model in which the production of relativistic particles and heating of dust in galaxy interactions gives rise to the observed radio-infrared correlation. The observations on which we report here were designed to test the radio-infrared correlation at the most luminous end by selecting luminous sources from a catalog of bright *IRAS* galaxies.

## *1.2 Radio Morphology*

The expected microwave emission from these galaxies in most models involves two components which have different morphologies. On the one hand, the microwave emission may be produced in a strongly nucleated starburst or an active nucleus, in which case we expect our maps to show an unresolved core of emission, perhaps with extended structures such as the small-scale jets or radio lobes seen in some low-luminosity Seyfert galaxies (Ulvestad & Wilson 1984; cf., Beichman *et al.* 1985 for a dramatic example of a double-lobed radio source associated with an *IRAS* galaxy). On the other hand, the microwave emission may be associated with star formation in the galaxy disk which will produce more diffuse extended emission, either via synchrotron emission from supernovae as modeled by Helou & Bicay (1993) or by bremsstrahlung from HII regions. The model of Harwit *et al.* (1987) proposes a third component, namely extended radio emission from relativistic particles accelerated in galaxy interactions. Condon *et al.* (1991c) suggest that values of $q$ (Eq. 1) can sometimes reveal sources with a nuclear component. Galaxies with $q < 2.0$ have excess radio emission above that correlated with the FIR; some of the excess radio emission may come from an active nucleus, rather than from the extended disk.

Condon *et al.* (1991c) give a simple but elegant argument that provides a useful diagnostic of non-thermal nuclear activity based on angular size measurements. Since the maximum FIR brightness temperature of an *IRAS* galaxy is the thermodynamic temperature of its emitting dust, which is typically $< 80^\circ$ K, an observed FIR flux translates directly to a minimum angular diameter of the star formation region. A galaxy with flux density $S_{60\mu\mathrm{m}}$ thus has a FIR-emitting region of *minimum* angular size $\approx 0.''07\,(S_{60\mu\mathrm{m}}/1\mathrm{Jy})^{\frac{1}{2}}$. The galaxies in our sample all have $S_{60\mu\mathrm{m}} > 2\,\mathrm{Jy}$, meaning that their *minimum* angular sizes are $> 0.''1$. If we assume that the radio and far-infrared emission are everywhere coincident in these galaxies, high-resolution radio maps have the potential to distinguish morphologically between thermal and non-thermal emission, although $0.''1$ is difficult to resolve even with the VLA in A array. However, a few of the sources in Condon *et al.* (1991c) have been detected with very long baseline interferometry (VLBI) on milli-arcsecond scales (Lonsdale, Smith, & Lonsdale 1993), implying that there must be an active nucleus in these systems by Condon *et al.* 's argument, and Norris, Allen, & Roche (1988a) and Norris *et al.* (1988b; 1990) have argued from their high-resolution radio studies that active galaxies are associated with unresolved radio cores.

We present here high resolution radio maps of a number of ultraluminous infrared galaxies from our VLA A-array observations. Our A-array maps have a beam size of $\approx 0.''5$ and resolve the extended emission in many of our sources. The resolution is unfortunately not high enough to apply the argument of Condon *et al.* , although the fact that we do see extended radio emission in many of these sources argues against the AGN hypothesis. Further argument and discussion are presented in §3.

### *1.3 Spectral Indices*

The radio emission from AGN results from the synchrotron process and is described by a decreasing power-law spectrum, i.e. $S_\nu \propto \nu^\alpha$ where $\alpha < 0$. As with most synchrotron sources, many AGN have spectral indices, $\alpha \approx -0.7$. However, compact radio sources tend to have flatter spectra, with $\alpha > -0.5$. We do not have the angular resolution to distinguish between compact and diffuse AGN at the typical distances of our sources, but of course our data are adequate to distinguish spectral indices of $-0.7$ from $-0.5$.

The more extended microwave emission associated with star formation can arise from three mechanisms, namely:

a. The young O and B stars in a starburst produce HII regions, which appear at microwave frequencies as bremsstrahlung sources. For the range of frequencies considered here, these sources have typical power-law spectra with $\alpha \approx -0.1$.

b. The most massive, short lived stars formed in a burst of star formation rapidly undergo supernova explosions, which in turn generate relativistic electrons, which produce synchrotron radiation. Here, the spectral index is typically $-0.6$ to $-0.8$, depending on the energy spectrum of the electrons.

c. The interstellar dust is heated by the visible and ultraviolet radiation from stars of a wide range of masses. The re-radiation by the dust produces copious FIR emission with a thermal spectrum (i.e., $S \propto \nu^{+2}$). The long wavelength tail of this radiation can appear in the millimeter wave region, although the bulk of it appears in the *IRAS* bands.

Thus microwave radiation samples different regions of the initial stellar mass function via these three mechanisms. The fact that microwave emission can arise from more than one mechanism makes even more remarkable the strikingly tight correlation between microwave fluxes and FIR fluxes in both normal and star-forming galaxies, and suggests that one mechanism dominates in most systems. Thus measurements of the spectral index of the microwave emission gives us clues about the star formation in these galaxies (Sramek & Weedman 1986). If the model of Harwit *et al.* (1987) is correct, however, then the radio emission from the sources in our sample does not probe star formation at all. In this case, the radio emission will be purely synchrotron radiation.

### *1.4 Outline*

The outline of the remainder of this paper is as follows: in the next section, we describe our sample selection, observations, and reductions. The contribution of this work in confirming relationship (1) at the high luminosity end appears in §3. The morphologies of our sources at the three resolutions afforded by our observations are discussed in §4 and our conclusions in §5.

## 2. Sample Selection, Observations, and Reductions

### 2.1 *Sample Selection*

Strauss *et al.* (1990; 1992) have carried out a redshift survey of *IRAS* galaxies, flux-limited to 1.936 Jy at 60$\mu$m and selected via infrared color criteria from the *IRAS* Point Source Catalog (1988). The sample covers 88% of the sky, and consists of 2658 galaxies. Although the median redshift of the sample is low, $\approx$ 4600 km s$^{-1}$, the power-law nature of the *IRAS* luminosity function at high luminosities (e.g., Soifer *et al.* 1987b; Saunders *et al.* 1990; Yahil *et al.* 1991) means that a flux-limited redshift survey contains a significant fraction of objects at very high luminosities. We define FIR luminosity in the standard way,

$$L_{\rm FIR} = 4\,\pi r^2 \nu S_{\rm FIR} = 117 \left(\frac{v}{h\ {\rm km\ s^{-1}}}\right)^2 \frac{S_{\rm FIR}}{1\,{\rm Jy}}\, L_\odot, \tag{3}$$

at frequency $\nu = 3.75 \times 10^{12}$ Hz (corresponding to $\lambda = 80\,\mu$m)[7]. The Strauss *et al.* sample contains 188 galaxies with FIR luminosities greater than $10^{11.15} h^{-2}\, L_\odot$ (or $10^{11.4}\, L_\odot$ with $h = 0.75$, as we adopt here). Here, $h$ is the Hubble Constant in units of $100\,{\rm km\,s^{-1}\,Mpc^{-1}}$. One hundred twenty-six of these ultraluminous galaxies are observable with the VLA (*i.e.*, $\delta > -20^\circ$). We selected 43 of these according to which areas of the sky were visible during our scheduled time, avoiding well-known galaxies which already had been observed extensively with the VLA. Our sample is listed in Table 1, which includes *IRAS* Point Source Catalog positions, fluxes in the 60 and 100$\mu$m bands, heliocentric redshifts, and the log of the FIR luminosities as given by Eq. (3).

Four of the sources, represented by asterisks in Table 1, were undetected with the VLA. All four of these are at low Galactic latitudes, and three have 60 to 100$\mu$m flux ratios typical of infrared cirrus. These were presumably sources in which an optical redshift was obtained of a background galaxy in the field unrelated to the *IRAS* emission, and thus the infrared luminosity was spurious. Indeed, in this case, the elevated $S_{100\,\mu{\rm m}}$ artificially boosts $S_{FIR}$, making such sources more likely to appear in our ultraluminous list. We will not discuss these sources further in this paper.

### 2.2 *Observations and Data Reduction*

We observed with the VLA in C array at 6 and 20 cm for a total of 17 hours on 1989 July 31 and August 3 and 4 in generally good weather. Center frequencies were 4860 and 1460 MHz. Typical exposures were 5 to 6 minutes per object at 6 cm, and 10 to 11 minutes per object at 20 cm. These integration times were chosen to produce roughly comparable sensitivities to sources with $\alpha = -0.7$ at our two observing frequencies. Phase errors were minimized by observing sources with similar sky positions in groups of three or four with observations of nearby phase calibrators after every group. Flux calibration was performed with observations of 1328+307 (3C286) and 2005+403. The effective resolutions at 6 and 20 cm were $\approx 4''$ and $\approx 15''$, respectively.

[7] Note that we do not apply a K-correction here; all the galaxies in the sample have $z < 0.16$, and the K-corrections are small (cf., Fisher *et al.* 1992).

We obtained subsequent observations with the VLA in A array at 6 cm on 1990 May 6, for 15 hours, during which time we obtained snapshots of 24 sources detected in the C-array observations. The weather at the site was perfect for the A array observations, with no clouds or wind. Typical exposures were 25 to 26 minutes. The object exposures were interleaved with observations of phase calibrators and 1328+307 was again used as the primary flux calibrator. The effective resolution was $\approx 0.''5$.

The flagging, calibration, and imaging of the C array data was performed with AIPS at Haverford College during the summer and fall of 1991 and 1992. The flagging and calibration of the A array data was performed at the VLA Array Operations Center in May, 1990, while the imaging was done at Haverford College during the summer and fall of 1991.

Flagging of bad data was performed with the AIPS task TVFLGR. Although the overall quality of the data was good, there were a few bad scans and some antennas were especially noisy. There were also assorted noise spikes which were flagged and removed.

Standard AIPS programs were used to calibrate the data. We employed the Baars *et al.* (1977) flux scale for our primary calibrator, 1328+307, whose fluxes at the center frequencies of our observations were taken to be 14.66 and 7.31 Jy in the L and C bands, respectively.

### 2.3 *Forming Images*

We mapped our sources using the standard AIPS task MX, which performs the two-dimensional Fourier Transform from the *uv* plane to the image plane and deconvolution of the beam using the CLEAN algorithm (Högbom 1974). We used a pixel size of $4''$ for the 20-cm observations, and $1.''2$ and $0.''1$ for the 6-cm observations in the C and A arrays, respectively, and we used natural weighting throughout. We cleaned each image lightly, then determined a value for the rms noise level in regions free of sources. Typical $1\sigma$ noise levels in our lightly cleaned maps were $40 - 80$ $\mu$Jy at 6 cm in C array, $200 - 500$ $\mu$Jy at 20 cm in C array, and $25 - 50$ $\mu$Jy at 6 cm in A array, depending on the integration time and the presence of bright sources in or near the area mapped. We identified all the sources in the field of view from these preliminary images, and then cleaned more deeply in boxes placed around the sources. In a handful of cases, in which the initial images showed apparent phase calibration errors, self-calibration was used to improve the image quality.

### 2.4 *Determination of Flux Densities*

Most of our radio maps contain only unresolved or barely resolved components. Other sources are composed of multiple components. In all these cases, determination of flux densities was fairly straightforward—we used the standard AIPS program IMFIT, which fits a two-dimensional Gaussian to each component to find estimates of the peak intensity, integrated intensity, angular size (or upper limits thereof) and position. For clearly unresolved sources, the peak and integrated fluxes were generally the same to within $\sim 6\%$. In the few fields in which objects were more than $30''$ (at 6 cm) or $90''$ (at 20 cm) from the field center, we corrected the fluxes for the primary beam response of the VLA. The correction was no larger than 2% in any case.

The results of the 6 cm observations are shown in Table 2. For each source, we list the centroids of the radio emission (epoch B1950) of each component, as well as the peak and integrated fluxes (in mJy). The positions are accurate to $\sim 0\farcs1$. Note that the C and A array data are listed on separate lines. Detected background sources clearly not associated with the *IRAS* galaxy are not listed. We also provide a key to the source morphology. Unresolved sources (at resolution $\sim 4''$ for the C array observations and $\sim 0\farcs5$ for the A array observations) are labeled "u" in column 6. Double and multiple sources are labeled "d" and "m," respectively (most of these sources are shown in one of the figures). Finally, "r" indicates sources that are resolved.

Two sources, 0415+175 and 0901+144, display structure that is not well described by a Gaussian, in which case integrated fluxes were estimated by summing the flux densities within the boundaries of the features. These two sources are indicated by asterisks in Table 2; we report no peak intensities for these sources.

### 2.5 *Separation of Nuclear and Extended Emission*

For those sources observed at 6 cm in both the C array and the A array, we can estimate the extended flux densities by subtracting the A-array *peak* flux from the C array *integrated* flux densities. The peak flux density in an A array map is that arising from the central $0\farcs5$ (corresponding to a linear scale of 300–800 $h^{-1}$ pc in the redshift range of these sources), while the integrated flux density in a C array map comes from a solid angle nearly a hundred times larger. In §3, we compare the nuclear, extended, and total flux densities of these objects with the FIR emission from these objects and test the microwave/FIR correlation of Equation 1 for each case. Note that the A array observations were made 9 months later than were the C array observations. Any temporal variation in the sources (which of course are ubiquitous for compact sources) will thus introduce unknown errors into these extended fluxes.

### 2.6 *Spectral Indices*

All our sources were well detected at both 6 and 20 cm in our C array maps, allowing us to compute the spectral index $\alpha$ in all cases. Table 3 lists our peak and integrated fluxes at 20 cm, together with spectral indices from the integrated fluxes, and the quantity $q$ as defined by Equation 1. For those sources with multiple components resolved in the 6 cm observations but unresolved at 20 cm, we compared the sum of the fluxes of the 6 cm sources with the 20 cm flux for the blended image. Since the FWHM of the beam at 20 cm is $\approx 3$ times larger than at 6 cm, there may be sources with diffuse emission that is resolved out at 6 cm, but detected at 20 cm, thereby causing us to underestimate our values of $\alpha$.

We were able to measure spectral indices for 40 objects (2 components of 1909+450 and 1 component for each of the other 38 sources); they ranged from $-1.13$ to $+2.13$, with only one source, 0524+010, displaying an inverted spectrum. We discuss this source in more detail in §3.2 below. Excluding 0524+010 and 1909+450(a), which is unlikely to be associated with the *IRAS* source, the mean spectral index is $-0.65$ with a range from $-0.21 \pm 0.09$ to $-0.99 \pm 0.03$. Thus most of these sources appear to be ordinary synchrotron emitters. The eight sources with $\alpha > -0.5$, though, might be compact objects with spectra affected by free-free absorption (Condon *et al.* 1991c).

## 3. Radio-FIR Correlation

We now address the issue of whether the radio flux densities of these ultraluminous *IRAS* sources fit the linear relationship given by Equation 1.

### 3.1 *Radio Observations from the Literature*

A number of groups and individuals have carried out observing programs with the VLA that parallel ours (e.g., Unger *et al.* 1989; Condon *et al.* 1990, 1991b, 1991c; Neff & Hutchings 1992; Sopp & Alexander 1991b, 1992; Stine 1992). Their samples were defined in various ways, but contain a number of ultraluminous *IRAS* sources for which we did not obtain data. Table 4 lists all relevant radio data for galaxies in our sample of 188 ultraluminous *IRAS* galaxies which we could glean from the literature. We list the most familiar name for each object in column 1. Column 2 contains the source of the radio data. Columns 3 and 4 are the radio positions in B1950 coordinates. Columns 5 and 6 contain the radio fluxes at 1460 and 4860 MHz. Most of these observations were made at slightly different frequencies; we corrected to our central frequencies using observed spectral indices where available, otherwise we assumed $\alpha = -0.7$. If several values of the flux density for a single source were available, we used the most recent or most precise. Columns 7 and 8 include the 60$\mu$m and 100$\mu$m flux densities for these sources in Jy. The heliocentric redshift is listed in column 9, and the FIR luminosity (again assuming $h = 0.75$) is listed in column 10. Unlike our main sample listed in Table 1, this list contains many well-known galaxies (and a few quasars!); moreover, the median redshift is appreciably lower.

Combining this sample with that in Table 1 gives us 63 objects with radio data, which is 1/3 of the complete ultraluminous sample. It is not obvious that this is a completely unbiased subset. As noted above, the objects we observed ourselves avoided well-known galaxies for which radio data exist in the literature (e.g., Arp 220 and 3C 273, which appear in Table 4). The far-infrared luminosity distribution of the combined sample matches closely that of the complete set of ultraluminous galaxies.

It is very important to bear in mind that many of the programs cited in Table 4 were carried out at different angular resolutions from those in our observations. In particular, the flux densities reported by Condon *et al.* (1982; 1991c), Neff & Hutchings (1992) and Stine (1992) were obtained with the VLA in its A configuration, and hence at higher resolution than our C array work. As a consequence, they may have resolved out some of the radio flux. The footnote to Table 4 provides related information for all references.

### 3.2 *The Radio-FIR Correlation*

Figure 1a shows the correlation of flux density at 1460 MHz with FIR flux for the combined sample. Objects known to be Seyfert galaxies or QSOs from their optical spectra are indicated with solid points; all other points either have HII region spectra, or do not have optical spectra in the literature. A strong correlation is indeed seen; the correlation coefficient of all the points except the two obvious outliers is 0.82 (as shown at the top of the plot). The two outliers are both found in Table 4: 3C 273 is of course a radio quasar, while 1345+125 = 4C +12.50 is a well-known powerful radio galaxy. The corresponding plot at 4860 MHz is shown in Figure 1b. Again, the correlation is strong. Also, there is no obvious segregation between those galaxies with optical spectra indicative of AGN and the others.

Figures 1c and 1d show the correlation between far-infrared and radio *luminosity*. At 20 cm, the luminosity is given by:

$$L(20\,\mathrm{cm}) = 4.6 \times 10^{-5} \left(\frac{v}{h \ \mathrm{km\ s^{-1}}}\right)^2 \left(\frac{S_{20\,\mathrm{cm}}}{1\,\mathrm{mJy}}\right) L_{\odot}. \tag{4}$$

The 6 cm luminosity is defined similarly, with a coefficient of $1.53 \times 10^{-4}$. 3C 273 and 4C +12.50 have not been included in panels c and d. The correlation coefficient in these two panels is significantly lower than in the upper two panels, simply because of the smaller range of the variables plotted.

We now restrict ourselves to those sources that we observed (Tables 2 and 3). Figure 2a shows the same correlation as Figure 1c for the 39 sources we observed with the VLA. Figure 2b is the same correlation as in Figure 1c for our 6-cm luminosities.

The mean $q$, as defined in Equation 1, of the combined sample in Figure 1c is 2.30 with a standard deviation $\sigma_q$ of 0.63. However, if we throw out the two outliers, we find $\langle q \rangle = 2.40$ with $\sigma_q = 0.26$ for 57 objects, in reasonable agreement with the value of 2.34 quoted by Condon *et al.* (1991c) for a sample extending to much lower far-infrared luminosities than ours. Condon *et al.* (1991c) quote a somewhat smaller scatter, $\sigma_q = 0.19$, but some of our higher scatter may be due to the heterogeneity of the sample we are using; we cannot guarantee that all the fluxes taken from the literature were calculated in the same way as ours. Indeed, the scatter in Figure 2a is 0.22 and the mean value of $q$ is 2.36, both in excellent agreement with the values quoted by Condon *et al.*

The previously observed correlations between far-infrared emission and various measures of star formation such as optical emission-line strength and near-infrared CO absorption (see §1) imply that star formation is the energy source responsible for the *IRAS* emission in lower luminosity galaxies. Thus the fact that we find an identical FIR-radio correlation in our sample strongly suggests that the far-infrared emission from the high luminosity sources is also due to star formation.

The mean value of $q$ for the 6-cm data plotted in Figure 1d, defined in analogy to Equation 1, yields $\langle q \rangle = 2.72$ with $\sigma_q = 0.18$ for 51 objects, after excluding the outliers; the distribution is even tighter than at 20 cm. Figure 2b shows a similar distribution; excluding the one outlier, 0524 + 010, this plot yields $\langle q \rangle = 2.71$ with $\sigma_q = 0.18$. This is in excellent agreement with the values $\langle q \rangle = 2.75$ and $\sigma_q = 0.14$ found by Condon & Broderick (1991) at 4850 MHz for 122 sources over a large range of far-infrared luminosity.

The fact that we see the same correlation at 6 and 20 cm for the sources that we observed, shown in Figures 2a and 2b, is no coincidence; it follows from the fact that the range of radio spectral indices for our sources is not large. Indeed, the exception to this is the source 0524 +010, which is the outlier represented by the open symbol in Figure 1b. Because of its unusual spectrum, 0524+010 cannot satisfy the correlation at both wavelengths. It is interesting, and probably coincidental, that this source fits the correlation very well at 20 cm, the wavelength at which the correlation was originally defined.

Let us now investigate the possibility that the FIR–radio correlation is due to star formation spread throughout the disks of these galaxies. Then the correlation would be tighter between FIR and *extended* radio emission than between FIR and total radio emission, since any additional radio emission from an active nucleus would add to the scatter in the relationship (and lower $q$). Figure 2c shows the luminosity correlation with the nuclear component of the radio emission subtracted out, leaving only the extended component. For the 24 sources for which we have A-array maps, we define the 6 cm extended luminosity as the difference of the C-array *integrated* luminosity and the A-array *peak* luminosity. In contradiction to our expectation, the resulting correlation with far-infrared luminosity is quite weak. Excluding 0524 +010, we find a correlation coefficient for this plot of only 0.18, and a mean $q$ of 3.14 with a scatter of 0.28. The subtraction of the measured nuclear flux from the total flux will add noise, but not enough to explain the observed increase in scatter. We thus conclude that it is *not* simply extended star formation which gives rise to the FIR-radio correlation. To test the importance of the *nuclear* component of the radio emission, we have also plotted the far-infrared luminosity versus the *peak* A-array luminosity at 6 cm in Figure 2d. In this plot, we have included the A array measurements of Neff & Hutchings (1992). Here we find $\langle q \rangle = 3.01$ and $\sigma_q = 0.37$. The correlation coefficient is high, 0.66, but the scatter in $q$ is twice that in Figure 2b. One is therefore forced to accept that the FIR-radio correlation is dependent on activity in *both* the disks and the nuclei of these galaxies. Since the FIR-radio correlation that we find is identical to that found at lower luminosities, where star formation is believed to be the dominant process, we infer that our sources have starburst activity in both the disks and nuclei.

We next explore correlations involving the spectral indices. There are three independent spectral indices that we can form with our data: the quantity $q$ defined at 20 cm, the radio spectral index between 20 and 6 cm, $\alpha$, given in Table 3, and the 60-100 $\mu$m spectral index, $\alpha_{IRAS}$. In Figure 3a we plot the far-infrared luminosities versus the FIR spectral indices, including the sources from Table 4. We observe only a very weak correlation. As has been noted by other authors (e.g., Fisher *et al.* 1992 and references therein), more luminous *IRAS* galaxies tend to have excess 60$\mu$m emission relative to their 100$\mu$m emission. A similar correlation is seen between the far-infrared luminosity and the radio spectral index (excluding 0524+010): the more FIR luminous sources tend to have flatter radio spectra. This is consistent with the argument by Condon *et al.* (1991c) that free-free opacity gives a relation between FIR brightness temperature (and therefore luminosity) and radio spectral index. significant

Figure 3c reveals a correlation between $q$ and $\alpha_{IRAS}$. The correlation reported by Fisher *et al.* (1992) shows that $\alpha_{IRAS}$ tends to be large in high-luminosity sources, and high FIR luminosity can produce large values of $q$. On the other hand, large values of $\alpha_{IRAS}$ have also been used as a signature of nuclear activity (e.g., Heisler & Vader 1994). The correlation in Figure 3c shows, however, that the sources most likely *a priori* to be AGN by this argument do not in fact exhibit strong radio emission.

Given the correlations found thus far, it is not surprising that strong correlations also exist between $\alpha_{radio}$ and $\alpha_{IRAS}$ (not shown) and between $\alpha_{radio}$ and $q$ (Figure 3d).

## 4. Source Morphology

Of the 39 *IRAS* sources, 18 are either multiple and/or marginally resolved in our C band C array observations with a resolution of $\approx 4''$. We display some of these sources in Figure 4. This resolution corresponds to $2-7\,h^{-1}$ kpc for the redshift range of our sources. Thus almost half of these ultraluminous *IRAS* sources either contain multiple radio components, or have radio emission distributed over extended disks.

Of the 24 maps made at the higher resolution ($\theta \approx 0\farcs5$, corresponding to $300-800\,h^{-1}$pc) of the A array, 16 sources or components are resolved or multiple. Figure 5 shows the A array maps of the more interesting sources. Three of the resolved sources have diffuse extended structure (see Figure 5b). Two contain jet-like sources (see Figure 5c), indicating the presence of "central engines."

Almost all the sources have larger 6 cm flux densities in the C array observations than in the A array observations, indicating that much of the flux comes from scales larger than 2 to $3''$, to which the A-array observations are insensitive. Two sources ($1909+450$(b) and $0958+471$) appear unresolved in the A array observations, but are resolved with the C array, indicating that all of the flux on scales larger than $0\farcs5$ is extended over scales larger than $4''$.

The mean spectral index of those sources that are *unresolved* at $\approx 4''$, excluding $0524+010$, is $-0.60 \pm 0.04$, while the mean for the resolved sources is $-0.71 \pm 0.05$, only marginally different. Both cases are consistent with synchrotron radiation as the source of the radio emission.

With some of the sources that appear resolved in the 6 cm C array observations, it was not entirely clear whether low-level structure around the sources was real, or an artifact of side-lobes of the synthesized beam. We used a number of techniques to decide between the two possibilities. We experimented with cleaning to different levels and with self-calibrating the images. We also looked at the residual map left after the best two-dimensional Gaussian model was subtracted from the actual map. One of these sources, $0932+613$, is displayed in Figure 4. Of the seven questionable cases, two were deemed resolved, and five unresolved (including $0932+613$).

Finally, in three cases, $0415+175, 1909+450$, and $2150-062$, the lower resolution 20 cm observations picked up larger scale radio structure not seen at 6 cm. Not surprisingly, the spectral indices are relatively steep for all three.

### 4.1 *Discussion of Individual Sources*

The radio maps of most of those sources which showed interesting structure are shown in Figures 4 (6 cm, C array) and 5 (6 cm, A array). None of the sources show appreciable structure at 20 cm, although a few are marginally resolved.

**0015 +545**: This is a double source, with separation $4\farcs7$ (Figure 5b). Both components are unresolved at $0.''5$ resolution. The optical spectrum shows emission lines characteristic of an H II region.

**0026 +425**: The optical spectrum of this source shows it to be a LINER. In the C array (Figure 4), it shows an intriguing second component without an obvious optical counterpart, although this second source is resolved out in the A array map (Figure 5a).

**0248 +430**: The 6 cm and 20 cm maps are displayed in Figure 6. The optical field contains a QSO at redshift 1.310, and a galaxy with two nuclei $\approx 14''$ to the South-East.

This system has been studied in detail because absorption lines at the redshift of the galaxy are seen in the spectrum of the QSO (Womble *et al.* 1990; Sargent & Steidel 1990). The galaxy itself has a double Seyfert nucleus (Kollatschny *et al.* 1991). At radio wavelengths, the QSO is a VLA calibrator; at 6 and 20 cm we determine its flux density to be 1.41 Jy and 1.03 Jy, respectively. This 6 cm flux is in good agreement with that of Gregory & Condon (1991), but White & Becker (1992) quote a flux of 834 mJy at 1.40 GHz, or 848 mJy at our observing frequency of 1.46 GHz. Thus the radio emission from the QSOmay be variable at 20 cm. Indeed, Witzel *et al.* (1979) report 0.38 Jy at 20 cm. In our image, there is also a weak 6 cm radio source coincident with the optical position of the galaxy, with an integrated flux density of $12.8 \pm 2.3$ mJy (the large error results from the proximity to the QSO; Figure 6). The *IRAS* position is coincident with the galaxy, not the QSO, and is separated from it enough that the contamination of any infrared emission from the QSO to the *IRAS* flux should be small. Moreover, *IRAS* emission from quasars at this redshift tends to be very weak. Unfortunately, our synthesized beam size at 20 cm is comparable to the galaxy-QSO separation, and the galaxy and QSO are barely resolved (Figure 6). Our estimate for the galaxy flux density at 20 cm, obtained from a 2-component Gaussian fit to the image, with positions derived from the 6 cm map, is $15 - 25$ mJy. This value is roughly consistent with Equation 1 and the FIR flux density.

**0110 +732**: The radio emission (Figure 5b) is quite extended and diffuse, suggestive of emission from a disk. The optical spectrum is that of an HII region.

**0315 +422**: This source coincides with a close pair of Seyfert galaxies, although only a single radio source is detected (Figure 5a).

**0415 +175**: This source has a one-sided jet $\approx 1.''5$ long (Figure 5c). The lower resolution image at 6 cm is barely resolved (Figure 4) in approximately the direction of the jet, at position angle 75°. The optical spectrum is that of a Seyfert 2. There is also a weak source $12.''4$ SE of the central source.

**0423 +143**: An extended North-South ridge of emission, with substantial substructure is apparent in the A-array map (Figure 5b). The optical image is very faint, but is also extended North-South. The optical spectrum shows lines characteristic of a LINER.

**0524 +010**: We discuss this source, which has a highly unusual radio spectrum, at length in §4.2 below.

**0807 +050**: This source shows what appears to be a jet in the A-array map (Figure 5c). The optical image shows tidal tails, and the optical spectrum is that of an HII region.

**0901 +144**: This source, which is also known as Markarian 1224, UGC 4756, and IC 2431, is coincident with a compact group of four galaxies; the two radio sources seen in Figures 4 and 5b correspond to the two brightest of these galaxies. Condon, Frayer, & Broderick (1991b) have observed this source at 6 cm with the VLA in D array; they barely resolve it, and find a total flux of 31 mJy, in good agreement with our total flux of 33 mJy. This source has also been observed at the VLA at 6 cm and 20 cm by Stine (1992) in the B array and A array, respectively. As Figures 4 and 5b show, the source has two components, both resolved. Stine's 6 cm, B array, flux density measurement for component (a) is 16.87±0.81 mJy, in good agreement with our 6 cm, C array, flux density of 15.88±0.97 mJy. For component (b), however, we obtain a 6 cm flux density ≈40% lower than Stine's. Since his resolution is greater, he cannot be detecting extended, low surface brightness emission which we miss. We suggest that the core of component (b) is variable and that its flux density dropped by ≈ 40% between Stine's observations and ours. Support for this suggestion is provided by our 20 cm fluxes. In Stine's 20 cm, A array observations, which clearly resolve components a and b, the sum of the flux densities is 82±2 mJy, while we obtain only 68±5 mJy in our lower resolution, blended image.

**0932+613**: This is the well-studied interacting galaxy UGC 5101. The A-array map (Figure 5a) shows possible low-level structure around a strong core, while in the C-array map (Figure 4) the source is unresolved. This source has been studied in detail by Sanders *et al.* (1988) who found it to have a Seyfert 2 spectrum. Like many of the sources in this sample, this source shows a disturbed morphology in the optical, indicating a recent merger. See Carico *et al.* (1990) for a 2.2 $\mu$m image, which reveals an east-west extension. Another radio image is presented by Neff & Hutchings (1992), whose total 6 cm flux is in good agreement with ours. This source was also observed at 20 cm with the VLA in A array by Condon *et al.* (1990), who obtained a flux density consistent with ours. Finally, Lonsdale *et al.* (1993) have observed this source using VLBI at 18 cm and find an unresolved milli-arcsecond core of 28 mJy. Thus some of the nuclear radio emission of this source is almost certainly associated with an active nucleus. We note that $q$ for this source is low.

**0958+471**: A pair of galaxies consisting of one Seyfert 1 and one Seyfert 2 galaxy are found in this field. Each of these sources is detected in both our 6 cm and 20 cm maps, but the *IRAS* position is coincident with the brighter radio source. We include the radio flux of only the source that coincides with the *IRAS* source in our analysis.

**1056+248**: Armus, Heckman & Miley (1989; 1990) find the optical spectrum of this interacting pair of spirals to match that of an HII region. Neff & Hutchings (1992) measure a total flux of 20 mJy at 6 cm for this source, in good agreement with our measurements. Condon *et al.* (1990) measure 50 mJy for this source at 20 cm, again in good agreement with our data.

**1909+450**: A pair of radio sources (Figure 4) coincides with an interacting pair of galaxies, each of which shows a strong HII region emission line spectrum.

**2008−031**: This source, shown in Figure 5a, is detected by White & Becker (1992) at 20 cm with a flux of 130 mJy, more than twice the flux that we see. Their large beam, however, includes a bright background double-lobed source 0.′8 away, unassociated with the *IRAS* source.

**2249 −180**: This source has been studied in detail by Sanders *et al.* (1988) who show it to be a merger remnant with an HII region spectrum. The 2.2$\mu$m image of Carico *et al.* (1990) is extended East-West, unlike our radio images at comparable or better resolution. This source is included in the survey of Condon *et al.* (1990). They obtain a 20 cm flux density in the A array of 6.1 ± 3.0 mJy, in good agreement with our value at much lower resolution, suggesting that the source is not heavily resolved. Their flux density measurements at 20 and 3.6 cm yield a spectral index of $-0.41 \pm 0.30$, in agreement with our $\alpha$ from 20 to 6 cm of $-0.40 \pm 0.05$. Neff & Hutchings (1992) measure a 6 cm flux density of 3.8 mJy at 6 cm, in agreement with our measured value of $3.40 \pm 0.15$ mJy. Lonsdale *et al.* (1993) did not detect this source with VLBI at 18 cm.

**2336 +360**: Sopp & Alexander (1992) observed this source previously at the VLA and find a 6 cm flux density in agreement with ours, and a 20 cm flux density 25% lower than ours. However, their 20 cm measurements were done in A array and probably resolved out some of the flux. Our 6 cm A array image is shown in Figure 5a.

### 4.2 *An unusual, inverted spectrum source*

Source 0524 +010 has a highly unusual radio spectrum. Morphologically, the source is unresolved at either 4″ or 0.″5 resolution. Near infrared images at 2.2 $\mu$m show hints of a double nucleus with separation of $\approx 2''$; if so, only one of the nuclei is radio-bright.

Our C array observations, as indicated in Tables 2 and 3, yield flux densities of $187 \pm 4$ mJy at 6 cm and $14.4 \pm 1.2$ mJy at 20 cm. Other observations provide support for these flux density values. Our A array observation at 6 cm, which may resolve out some extended flux, is lower than our C array flux density, but is still well above that at 20 cm. Since the A array observation occurred several months later, some of the difference between the A and C array flux densities could be due to temporal variation. Gregory & Condon (1991), using a single dish, find a 6-cm flux density which lies between our C and A array values. The relatively low flux density at 20 cm is confirmed by a D array VLA snapshot observation made for us in September 1993 by Alex Rudolph. An observation at 3.6 cm by Alex Rudolph at the VLA also in September 1993 and a series of observations at 1.95 cm by Mike Jones at the Mullard Radio Astronomy Observatory from mid-March 1994 to late May 1994 yield flux densities of $130 \pm 3$ mJy and $68 \pm 3$ mJy, respectively. Thus the flux of 0524 +010 peaks in the frequency range 5 − 8 GHz, and has a spectral index between −1.2 and −1.1 at higher frequencies. The Mullard fluxes also show no evidence of variability at 1.95 cm.

This object fits the classification of a "gigahertz peaked spectrum" (GPS) radio source. An extensive study of such sources and a discussion of the physical implications of their observed characteristics can be found in O'Dea, Baum, & Stanghellini (1991, hereafter referred to as OBS). However, this source has a flux (and luminosity) far below those of any of the sources studied by OBS. There is one other known GPS source in the Strauss *et al.* (1992) sample of ultraluminous *IRAS* galaxies, 1345 +125 = 4C +12.50. However, this source is 550 times more luminous than 0524 +010 at 20 cm.

OBS noted that there is only one known GPS source with redshift below 0.1. The much lower luminosity of 0524+010 brings up the intriguing possibility that there are a number of lower redshift GPS sources with much lower luminosities, in analogy with the Seyfert galaxy–quasar connection.

The radio spectral shape of 0524+010 is very similar to that of one of the sources in the sample of O'Dea *et al.* (1990), namely 0108 +388. The peak of the spectrum, around 6 GHz, low-frequency spectral index, of 2.1, and the high-frequency spectral index, around −1.1, all place this source at an extreme of the O'Dea *et al.* sample. The steep spectrum in the optically-thick realm is very close to the theoretical synchrotron self-absorption slope, and so suggests that this source, if its spectral shape is determined by synchrotron self-absorption, has an almost uniform structure in terms of density, electron energies, and magnetic field. It has been suggested, however, that the turnover at gigahertz frequencies of compact sources could also be due to free-free absorption by surrounding thermal gas (van Breugel 1984; Condon *et al.* 1991c). The sharpness of the turnover in 0524+010 does suggest that free-free absorption is present in this source (Condon, private communication). This premise is supported by the optical spectrum, which resembles that of HII regions; the central engine appears to be obscured by surrounding starburst gas. On the other hand, the radio spectrum of 0108+388 was found by Baum *et al.* (1990) to conflict with the exponential turnover expected with free-free absorption. Flux density measurements at a number of frequencies below the turnover would reveal whether the spectrum of 0524+010 fits the free-free absorption or synchrotron self-absorption model better.

The FIR spectrum of 0524 +010 peaks at 60$\mu$m, which is often associated with a hot FIR component arising from an AGN (Vader *et al.* 1993; Heisler & Vader 1994). There are several such "60$\mu$m peakers" in our sample. As mentioned above, the optical spectrum shows emission lines characteristic of star formation. However, any AGN may well be hidden at optical wavelengths by large quantities of dust. We are obtaining additional optical and near-infrared spectroscopy on this object and hope to determine its flux densities at 0.9 cm and 0.3 cm in the near future.

Finally, we note that the 20 cm to 6 cm spectral index of this source matches almost exactly that of the 2.73 K cosmic microwave background radiation (CBR). Thus the radio emission from a source like 0524 +010 would be spectrally indistinguishable from a fluctuation in the CBR. Do sources like 0524 +010 present a problem for searches for fluctuations in the CBR? Given that we found only one such source in an extreme sample of objects, we can not estimate how common such sources may be. We can say that 0524 + 010 would mimic a CBR hotspot of amplitude $\frac{\Delta T}{T} \sim 3 \times 10^{-4}(\Theta/1^\circ)^{-2}$, where $\Theta$ is the half-power beam width used, *if* the observations were made at $\lambda = 6$ cm. Fortunately, most searches for CBR fluctuations are made at shorter wavelengths, typically $\lambda \leq 1.5$ cm. At $\lambda = 1.5$ cm, the flux density of 0524 +010 has dropped to 68 mJy, corresponding to $\frac{\Delta T}{T} \sim 6 \times 10^{-5}(\Theta/1^\circ)^{-2}$. The real concern for CBR observations is the possibility that GPS sources exist with flux density peaks between 10 and 30 GHz rather than a few GHz. Consider, for instance, a source with a spectrum like 0524 +010 but peaking at 30 GHz at a flux density of 1 Jy. Such a source would produce a hotspot of amplitude $\frac{\Delta T}{T} \sim 1.3 \times 10^{-5}$ in 0.9 mm observations like those of Gaier *et al.* (1992; with $\Theta \sim 1.^\circ6$), yet be well below the threshold of lower frequency radio catalogs used by CBR observers to locate foreground, contaminating, radio sources. The problem would be even more severe if such sources were variable. It is worth noting explicitly that a source like the one hypothesized here would not have been included in the sample of O'Dea *et al.* (1990), since it would have been too weak at 2.7 or 5.0 GHz. Clearly some caution in the use of low frequency catalogs to correct CBR observations is in order. In this connection, note the very recent observations by Church *et al.* (1995) to look for possible foreground sources causing apparent non-Gaussian fluctuations in two recent reports of anisotropy in the CBR.

## 5. Discussion and Conclusions

We have made radio maps at 6 and 20 cm of a statistically complete sample of 39 ultraluminous *IRAS* galaxies in order to address the source of the far-infrared luminosity: AGN or starburst? We listed three potential flags of nuclear activity in the introduction.

1. The far-infrared–radio correlation. We have found that the tight correlation between far-infrared and 20 cm flux seen at lower luminosities is obeyed even at these extreme luminosities, with no discernible increase in scatter. Since star formation and thermal radiation from dust heated by the interstellar radiation field are believed to be the primary sources of far-infrared emission from the lower-luminosity systems for the reasons discussed in §1, the presence of the same far-infrared–radio correlation at high luminosities implies that the same physical processes give rise to the emission. Interestingly, the correlation is much worse with just the *extended* radio emission, as well as with just the *nuclear* component. The nature of the activity that causes this correlation, therefore, must occur both in the nuclei and the disks of the galaxies.
2. The radio morphology. We found convincing evidence ("smoking guns") for only a few AGN, all involving jet-like components and no double-lobe sources. In particular, 0415 +175 shows a nuclear jet, which may be related to the Seyfert 2 nucleus seen in the optical. Indeed, this source has the lowest value of $q$ in the sample, which does suggest a radio component unassociated with the far-infrared emission, as pointed out by Condon *et al.* (1991c). On the other hand, the other source with an apparent nuclear jet, 0807 +050, has a normal value of $q$ and an HII optical spectrum. We may also ask about the morphology of the 5 sources with the lowest values of $q$ ($q < 2.1$). One is 0415 +175; another is 0932 +613, or UGC 5101, discussed above; another, 1031 +350, is extended at $\Theta \sim 4''$ resolution; and the other two are unresolved.

Many of our sources are unresolved, although our resolution is not adequate, even with the A array at 6 cm, to be able to apply the analysis of Condon *et al.* (1991c) rigorously, and thus to use limits on the angular size to give limits on the brightness temperature.

It should be noted that several of the sources in the sample have milli-arcsec radio cores detected with VLBI (Lonsdale *et al.* 1993); these sources certainly have an active nucleus. However, even in these cases, the fact that the radio and far-infrared emission follow the same correlation strongly suggests that the bulk of the far-infrared emission is powered by star formation, not the active nucleus.

3. The radio spectral index. The majority of our sources have spectral indices indicative of synchrotron emission. There is one source, however, 0524 +010, which shows a strongly inverted spectrum. We have suggested that this is a member of a potential new class of objects, intrinsically weak gigahertz-peaked spectrum sources.

Are there AGN's in the classical sense among the ultraluminous *IRAS* galaxies? Of course there are; a number of sources in our sample show Seyfert 1 spectra in the optical. Indeed, there are several dramatic radio galaxies and QSO's in the ultraluminous *IRAS* sample, including 3C 273, whose far-infrared emission must be dominated by the AGN. However, the bulk of our galaxies show no indication in our radio data that their far-infrared emission is powered by anything other than a starburst, despite luminosities approaching that of quasars. Thus we agree with the conclusions of Condon *et al.* (1991c), which were based on the angular extent of his high-resolution radio images. The triggering mechanism of such a dramatic starburst remains a mystery.

**Acknowledgements:** We would like to thank Gordon Kraus, who helped with the observations, and Jennifer Batty, Eric Richards, Leslie Sherman, and especially Scott Harrison who assisted in the analysis of the data. We are indebted to Alex Rudolph and Mike Jones for their observations of 0524 +010. This research has made use of the NASA/IPAC Extragalactic Database (NED) which is operated by the Jet Propulsion Laboratory, Caltech, under contract with the National Aeronautics and Space Administration. MAS is supported at the IAS by the W.M. Keck Foundation; work at Haverford College was supported in part by grant AST89-14988 from the National Science Foundation and by the Keck Northeast Astronomy Consortium which is funded by the W. M. Keck Foundation.

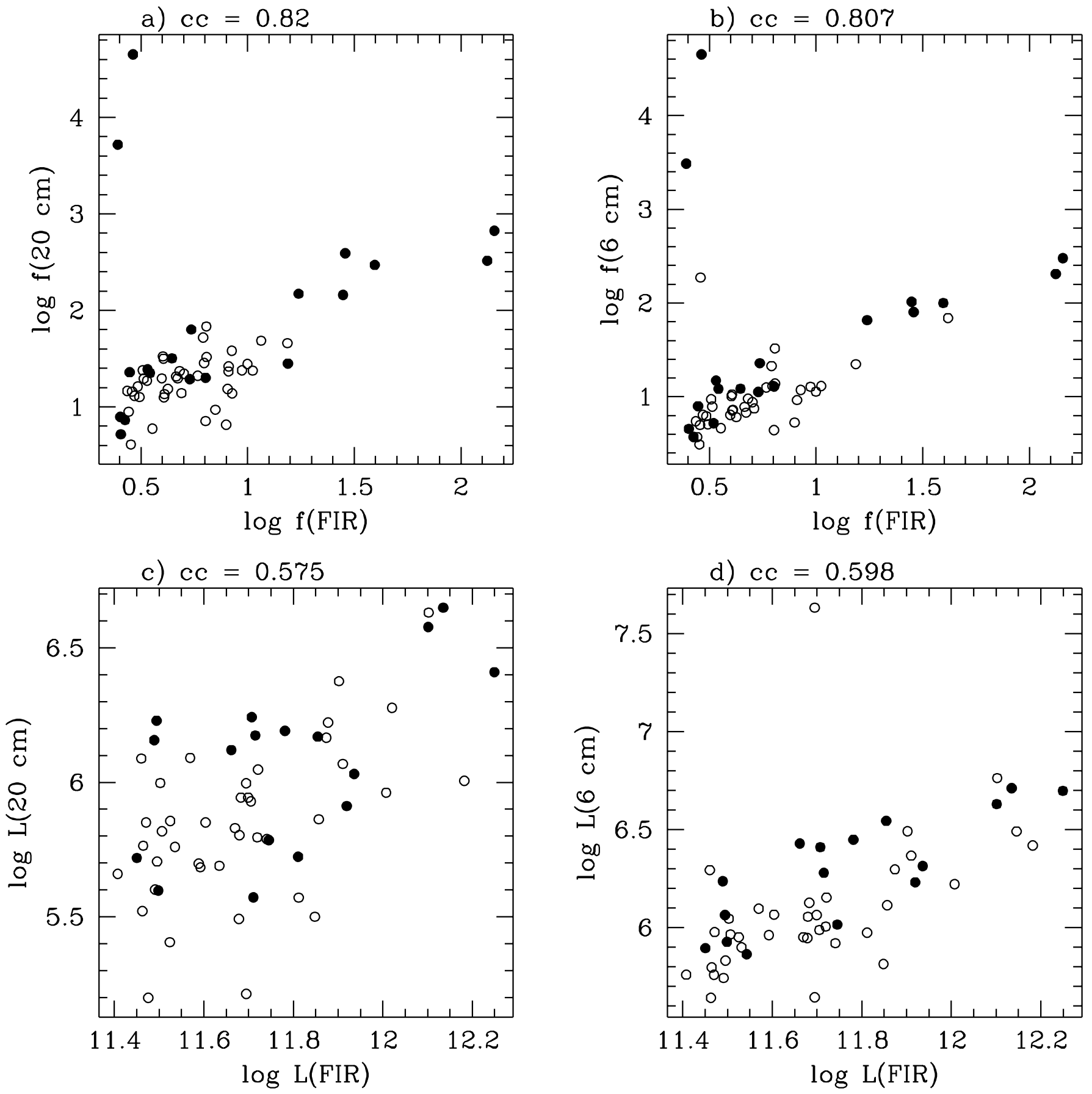

**Figure 1**: FIR–radio correlation for a combination of sources in our sample and sources drawn from the literature.

a. Flux density at 1460 MHz plotted against flux density at 80 $\mu$m. Solid points are objects whose optical spectra show evidence of an AGN while open points either do not have available optical spectra or show emission lines characteristic of an HII region. The two dramatic outliers are 3C 273 and 4C +12.50.

b. Flux density at 4860 MHz plotted against flux density at 80 $\mu$m.

c. Luminosity at 1460 MHz plotted against luminosity at 80 $\mu$m. 3C 273 and 4C +12.50 are off-scale.

d. Luminosity at 4860 MHz plotted against luminosity at 80 $\mu$m. 3C 273 and 4C +12.50 are off-scale; the point with log L = 7.6 is 0524 + 010 (see text).

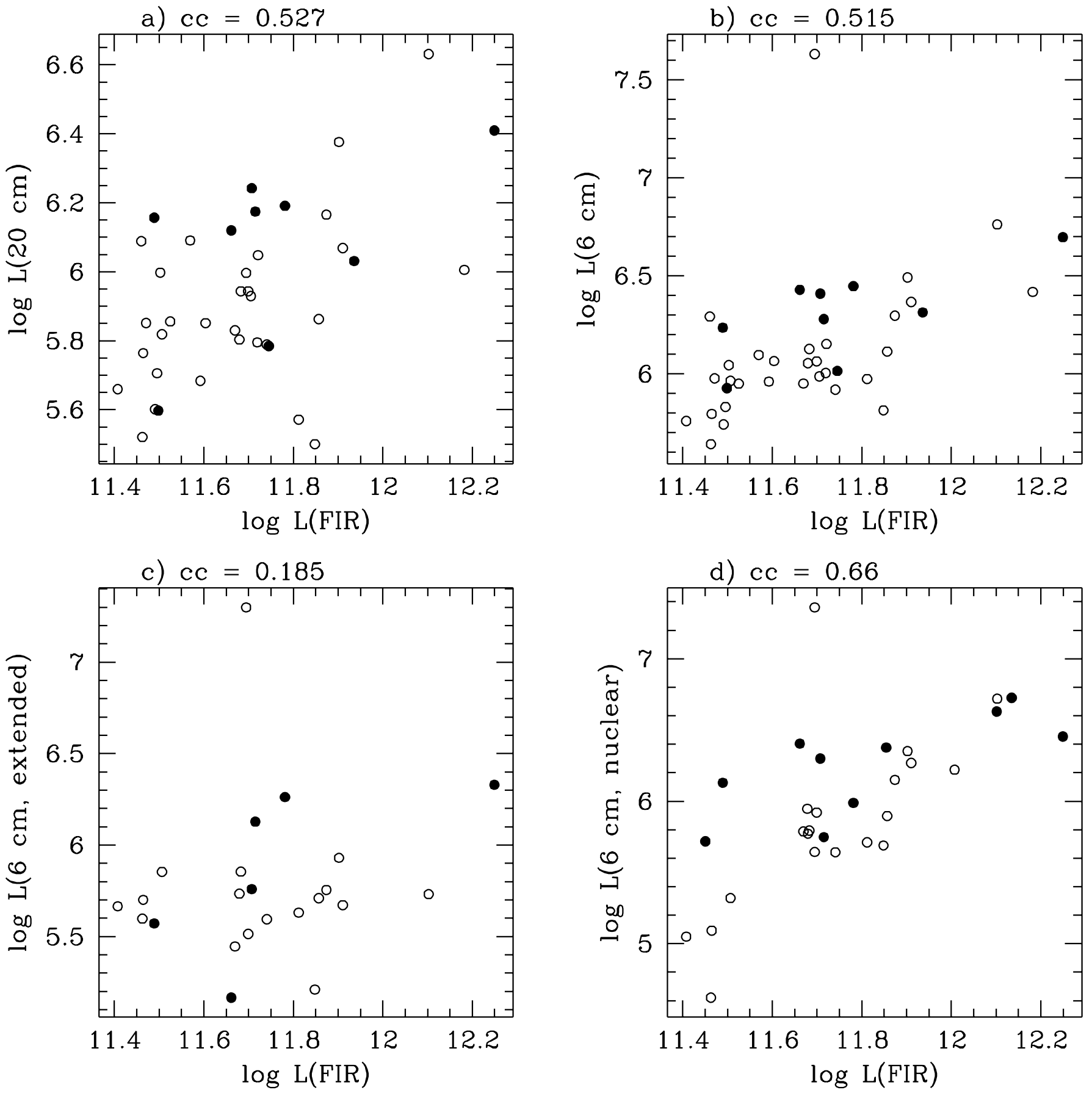

**Figure 2**: FIR–radio correlation for sources in our sample.

a. Luminosity at 1460 MHz plotted against luminosity at 80 $\mu$m.

b. Luminosity at 4860 MHz plotted against luminosity at 80 $\mu$m.

c. *Extended* luminosity at 4860 MHz plotted against luminosity at 80 $\mu$m (see text for details).

d. *Nuclear* luminosity at 4860 MHz plotted against luminosity at 80 $\mu$m (see text for details). Flux densities from Neff & Hutchings (1992) are included where available.

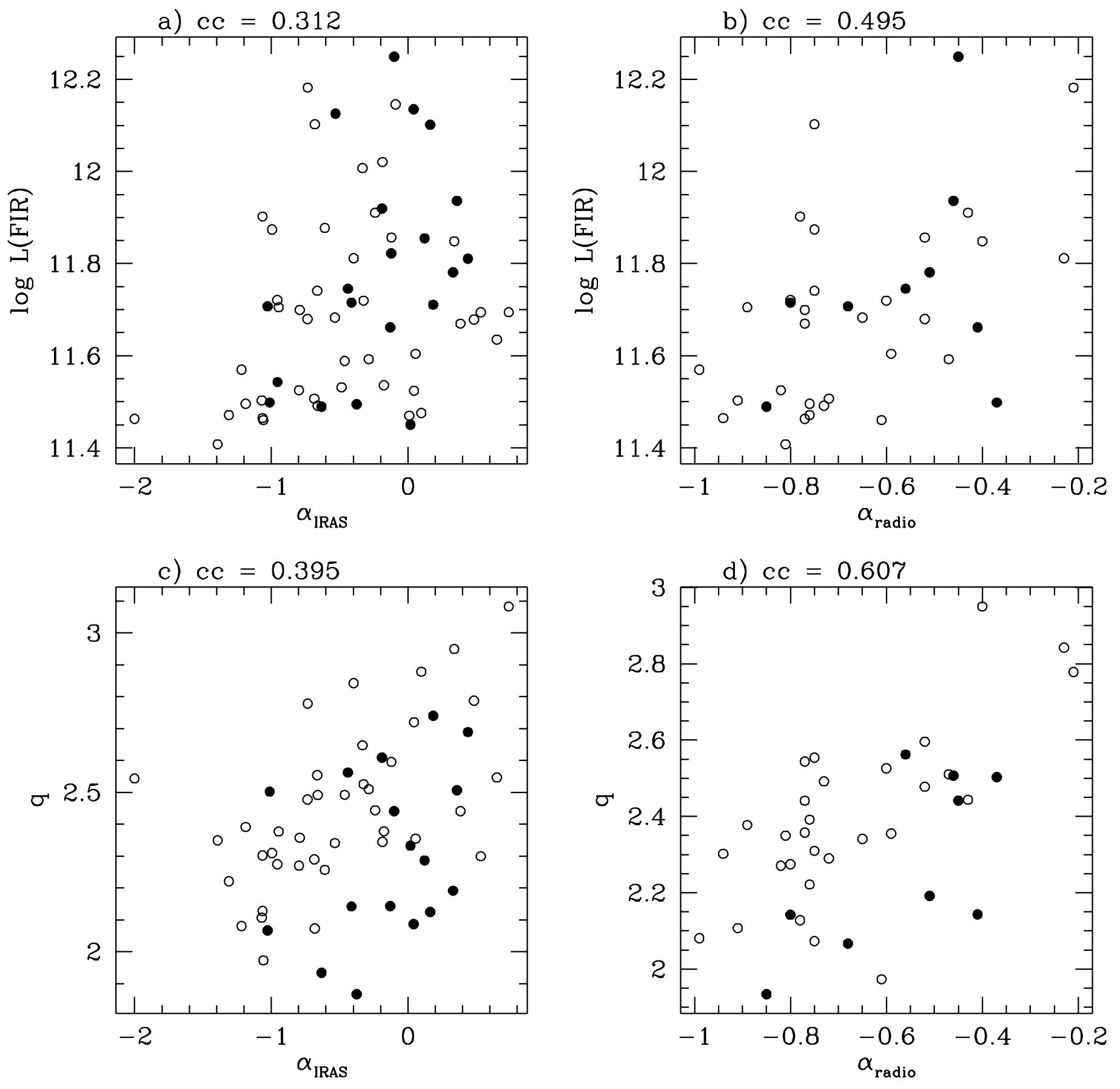

**Figure 3**: Correlation plots involving spectral indices.

a. Far-infrared luminosity vs. far-infrared spectral index.

b. Far-infrared luminosity vs. radio spectral index.

c. The value $q$ vs. far-infrared spectral index.

d. The value $q$ vs. radio spectral index.

*Note: Figures 4-6 are not available in postscript; please contact Margaret Best (best@ias.edu) for hardcopies.*

**Figure 4**: Contour plots of 6 cm flux at 4.″0 resolution. The sources displayed are $0026+425$, $0415+175$, $0901+144$, $0932+613$, $1019+132$, and $1909+450$. In this figure and all subsequent ones, the contours are generally drawn at the $\pm 3\sigma$, $\pm 5\sigma$, $7\sigma, 10\sigma, 15\sigma, 25\sigma, 40\sigma, 60\sigma, 85\sigma, 110\sigma, 150\sigma, 200\sigma$, and $300\sigma$ levels. The rms noise level, $\sigma$, is found in source-free regions for *each map separately.*

**Figure 5**: Contour plots of 6 cm flux at 0.″5 resolution.

a. Unresolved or barely resolved sources. The displayed sources are $0026+425$, $0315+422$, $0932+613$, $0958+471$, $1049+442$, $1836+354$, $2008-030$, and $2336+360$. Weak extended structure is visible in a few cases, such as $0932+613$ and $1049+442$. A few images appear to show resolved structure which is not significant because of a low signal-to-noise ratio (e.g., $0315+422$) or a shape similar to that of the dirty beam (e.g., $0958+471$).

b. Extended or multiple sources. Shown are sources $0015+545, 0110+732, 0423+143$, and $0901+144$.

c. Two cases of arcsecond jets. The displayed sources are $0415+175$ and $0807+050$.

**Figure 6**: Contour plots of $0248+430$.

a. In the 6 cm, C array map the brighter source is a QSO (a VLA calibrator with 6-cm flux density of 1.03 Jy). The weaker source 14″ East of the QSO is coincident with the *IRAS* galaxy (see text for discussion).

b. At 20 cm, the two sources are blended.

## Table 1. IRAS Properties of Sample

| RA (1950) | Dec | $S_{60\mu m}$ | $S_{100\mu m}$ | $cz$ | $\log \frac{L_{FIR}}{L_\odot}$ |
|---|---|---|---|---|---|
| | | Jy | Jy | km s$^{-1}$ | |
| (1) | (2) | (3) | (4) | (5) | (6) |
| 00 15 23.0 | +54 54 37 | 2.22 | 2.72 | 33448 | 11.81 |
| 00 26 13.9 | +42 51 45 | 3.03 | 2.56 | 29129 | 11.78 |
| 01 10 41.5 | +73 22 30 | 4.52 | 12.56 | 13217 | 11.46 |
| 01 52 09.8 | +52 24 45 | 3.01 | 4.38 | 23957 | 11.68 |
| 02 48 20.4 | +43 02 56 | 4.45 | 7.46 | 15571 | 11.50 |
| *02 52 53.1 | +43 50 46 | 3.00 | 2.22 | 33678 | 11.89 |
| 03 15 51.2 | +42 27 30 | 4.39 | 4.62 | 40288 | 12.25 |
| 03 28 49.0 | +01 08 12 | 2.47 | 2.40 | 25840 | 11.60 |
| *03 45 14.7 | +23 20 36 | 4.17 | 27.04 | 24905 | 12.21 |
| 03 50 19.3 | −03 28 55 | 3.23 | 3.74 | 21730 | 11.59 |
| 03 52 08.5 | +00 28 20 | 2.64 | 3.84 | 45622 | 12.18 |
| 04 15 28.3 | +17 55 23 | 4.09 | 5.65 | 16659 | 11.49 |
| 04 23 15.2 | +14 36 53 | 3.45 | 4.26 | 23972 | 11.72 |
| *05 15 32.7 | +06 36 47 | 6.22 | 38.02 | 8734 | 11.45 |
| 05 24 41.5 | +01 03 32 | 2.56 | 1.95 | 29037 | 11.69 |
| 07 38 44.7 | −03 44 36 | 2.98 | 6.07 | 16455 | 11.41 |
| 08 07 08.7 | +05 09 59 | 4.78 | 6.78 | 15650 | 11.51 |
| 08 34 05.4 | +15 50 43 | 2.07 | 2.90 | 23414 | 11.49 |
| 08 57 59.3 | +34 47 14 | 2.79 | 4.82 | 19645 | 11.50 |
| 09 01 48.8 | +14 47 38 | 4.44 | 7.62 | 14847 | 11.46 |
| 09 06 11.7 | −12 48 48 | 3.67 | 5.50 | 22073 | 11.70 |
| 09 11 11.1 | −10 07 01 | 7.46 | 10.47 | 16438 | 11.74 |
| 09 32 04.7 | +61 34 37 | 12.08 | 20.42 | 12000 | 11.71 |
| 09 58 21.7 | +47 14 09 | 2.77 | 2.96 | 25717 | 11.66 |
| 10 03 35.5 | +48 52 25 | 4.78 | 6.28 | 19421 | 11.68 |
| 10 19 01.4 | +13 22 04 | 3.33 | 5.40 | 22987 | 11.70 |
| 10 31 11.1 | +35 07 38 | 2.69 | 5.01 | 21274 | 11.57 |
| 10 37 49.1 | +11 09 08 | 2.21 | 1.84 | 40843 | 11.94 |
| 10 49 30.0 | +44 24 46 | 3.37 | 5.60 | 27616 | 11.87 |
| 10 56 35.4 | +24 48 43 | 12.16 | 14.35 | 12926 | 11.72 |
| 11 06 56.9 | +27 11 22 | 2.14 | 4.18 | 21080 | 11.47 |
| 18 36 49.5 | +35 49 36 | 2.23 | 3.84 | 34825 | 11.90 |
| 19 09 39.8 | +45 02 24 | 2.74 | 4.72 | 18980 | 11.46 |
| 20 08 46.4 | −03 08 52 | 4.61 | 6.53 | 31652 | 12.10 |
| 20 41 28.3 | −16 51 12 | 4.69 | 5.30 | 26107 | 11.91 |
| 21 39 37.3 | +36 23 22 | 2.16 | 3.52 | 29034 | 11.72 |
| 21 44 17.2 | +00 07 20 | 2.10 | 3.85 | 22187 | 11.50 |
| 21 50 28.4 | −06 28 55 | 3.70 | 3.04 | 23263 | 11.67 |
| *21 53 26.4 | +35 04 40 | 1.78 | 7.83 | 31111 | 11.92 |
| 22 08 20.4 | +27 03 35 | 1.99 | 2.99 | 24510 | 11.52 |
| 22 49 09.5 | −18 08 18 | 5.53 | 4.65 | 23312 | 11.85 |
| 23 32 42.7 | +29 13 25 | 2.07 | 2.59 | 31981 | 11.75 |
| 23 36 31.5 | +36 04 26 | 7.69 | 8.18 | 19338 | 11.86 |

* Objects undetected in radio. IRAS signal due to infrared cirrus.

**Table 2. 6 cm Radio Positions and Fluxes**

| Array | RA (1950) | Dec | $S_{6\,\mathrm{cm}}$ Peak mJy | $S_{6\,\mathrm{cm}}$ Int. mJy | |
|---|---|---|---|---|---|
| (1) | (2) | (3) | (4) | (5) | (6) |
| | | 0015+545 | | | |
| A | 00 15 23.07 | +54 54 36.6 | $1.26 \pm 0.02$ | $1.23 \pm 0.06$ | u |
| A | 00 15 23.24 | +54 54 32.2 | $0.43 \pm 0.01$ | $0.51 \pm 0.04$ | u |
| C | 00 15 23.14 | +54 54 35.2 | $2.03 \pm 0.06$ | $3.10 \pm 0.20$ | d |
| | | 0026+425 (a) | | | |
| A | 00 26 12.92 | +42 51 40.7 | $4.22 \pm 0.05$ | $4.67 \pm 0.13$ | r |
| C | 00 26 12.91 | +42 51 40.8 | $6.53 \pm 0.10$ | $8.02 \pm 0.28$ | r |
| | | 0026+425 (b) | | | |
| C | 00 26 13.22 | +42 51 48.3 | $1.59 \pm 0.11$ | $4.16 \pm 0.35$ | r |
| | | 0110+732 | | | |
| A | 01 10 41.65 | +73 22 22.6 | $0.88 \pm 0.02$ | $4.62 \pm 0.07$ | r |
| C | 01 10 41.54 | +73 22 22.9 | $5.72 \pm 0.09$ | $9.23 \pm 0.03$ | r |
| | | 0152+522 | | | |
| A | 01 52 09.30 | +52 24 42.2 | $3.79 \pm 0.03$ | $4.06 \pm 0.75$ | r |
| C | 01 52 09.32 | +52 24 42.5 | $5.89 \pm 0.08$ | $7.28 \pm 0.02$ | u |
| | | 0248+430 | | | |
| C | 02 48 20.00 | +43 02 52.5 | $10.59 \pm 0.73$ | $12.8 \pm 2.3$ | u |
| | | 0315+422 | | | |
| A | 03 15 51.67 | +42 27 33.2 | $6.45 \pm 0.14$ | $8.02 \pm 0.43$ | u |
| C | 03 15 51.67 | +42 27 33.3 | $10.72 \pm 0.07$ | $11.31 \pm 0.22$ | u |
| | | 0328+010 | | | |
| C | 03 28 49.17 | +01 08 14.0 | $6.34 \pm 0.11$ | $6.41 \pm 0.32$ | u |
| | | 0350−032 | | | |
| C | 03 50 19.11 | −03 28 55.2 | $6.99 \pm 0.09$ | $7.08 \pm 0.28$ | u |
| | | 0352+002 | | | |
| C | 03 52 08.01 | +00 28 17.2 | $4.39 \pm 0.06$ | $4.62 \pm 0.17$ | u |
| | | 0415+175 (a) | | | |
| A | | Jet | | 15.3* | r |
| C | 04 15 28.55 | +17 55 24.9 | $20.32 \pm 0.05$ | $21.33 \pm 0.15$ | r |
| | | 0415+175 (b) | | | |
| C | 04 15 28.94 | +17 55 14.6 | $1.11 \pm 0.03$ | $1.52 \pm 0.10$ | r |
| | | 0423+143 | | | |
| A | 04 23 15.12 | +14 36 52.6 | $2.63 \pm 0.06$ | $4.37 \pm 0.19$ | r,m |
| A | 04 23 15.12 | +14 36 54.5 | $0.73 \pm 0.02$ | $1.03 \pm 0.07$ | r |
| A | 04 23 15.13 | +14 36 56.3 | $0.22 \pm 0.01$ | $0.69 \pm 0.05$ | r |
| C | 04 23 15.12 | +14 36 53.3 | $11.73 \pm 0.04$ | $12.14 \pm 0.12$ | r |
| | | 0524+010 | | | |
| A | 05 24 41.33 | +01 03 31.8 | $100.00 \pm 0.74$ | $104.96 \pm 2.10$ | u |
| C | 05 24 41.34 | +01 03 31.6 | $185.76 \pm 1.32$ | $186.87 \pm 4.31$ | u |
| | | 0738−034 | | | |
| A | 07 38 44.65 | −03 44 39.6 | $1.52 \pm 0.02$ | $2.57 \pm 0.06$ | r |
| C | 07 38 44.93 | −03 44 40.1 | $5.61 \pm 0.13$ | $7.82 \pm 0.43$ | r |
| | | 0807+050 | | | |
| A | 08 07 08.42 | +05 10 01.5 | $3.14 \pm 0.10$ | $7.82 \pm 0.29$ | r |
| C | 08 07 08.44 | +05 10 01.4 | $11.97 \pm 0.14$ | $13.77 \pm 0.42$ | u |
| | | 0834+155 | | | |
| C | 08 34 05.40 | +15 50 46.6 | $3.54 \pm 0.07$ | $3.71 \pm 0.22$ | u |
| | | 0857+344 | | | |
| C | 08 57 59.30 | +34 47 13.4 | $9.92 \pm 0.17$ | $10.55 \pm 0.52$ | r |
| | | 0901+144 | | | |
| A | | Jet | | 4.44* | r |
| A | | Jet | | 3.42* | r |

**Table 2**—*Continued*

| Array | RA (1950) | Dec | $S_{6\,\mathrm{cm}}$ Peak mJy | $S_{6\,\mathrm{cm}}$ Int. mJy | |
|---|---|---|---|---|---|
| (1) | (2) | (3) | (4) | (5) | (6) |
| | | 0901+144 (a) | | | |
| C | 09 01 48.60 | +14 47 39.9 | 7.16 ± 0.41 | 16.89 ± 1.49 | r |
| | | 0901+144 (b) | | | |
| C | 09 01 49.03 | +14 47 35.7 | 8.97 ± 0.33 | 15.88 ± 0.97 | r |
| | | 0906−124 | | | |
| A | 09 06 11.47 | −12 48 47.9 | 5.76 ± 0.06 | 7.14 ± 0.18 | r,m |
| A | 09 06 11.59 | −12 48 43.0 | 0.52 ± 0.03 | 0.70 ± 0.08 | r,m |
| C | 09 06 11.51 | −12 48 44.4 | 7.43 ± 0.15 | 8.74 ± 0.48 | r |
| | | 0911−100 | | | |
| A | 09 11 10.76 | −10 07 04.2 | 5.96 ± 0.07 | 9.11 ± 0.20 | u |
| C | 09 11 10.80 | −10 07 05.3 | 9.93 ± 0.17 | 11.25 ± 0.51 | r |
| | | 0932+613 | | | |
| A | 09 32 04.77 | +61 34 36.9 | 50.96 ± 0.47 | 61.39 ± 1.42 | r |
| C | 09 32 04.61 | +61 34 38.0 | 57.94 ± 0.94 | 65.39 ± 2.90 | u |
| | | 0958+471 | | | |
| A | 09 58 21.50 | +47 14 14.2 | 14.10 ± 0.08 | 15.48 ± 0.25 | u |
| C | 09 58 21.47 | +47 14 14.0 | 14.77 ± 0.03 | 14.86 ± 0.08 | r |
| | | 1003+485 (a) | | | |
| A | 10 03 35.68 | +48 52 24.4 | 1.12 ± 0.06 | 2.41 ± 0.19 | u |
| C | 10 03 35.44 | +48 52 19.9 | 3.68 ± 0.13 | 4.76 ± 0.43 | u |
| | | 1003+485 (b) | | | |
| A | 10 03 36.07 | +48 52 28.9 | 4.95 ± 0.06 | 5.49 ± 0.18 | u |
| C | 10 03 35.97 | +48 52 28.4 | 7.77 ± 0.15 | 8.32 ± 0.43 | u |
| | | 1019+132 | | | |
| C | 10 19 02.36 | +13 22 04.0 | 5.01 ± 0.19 | 6.74 ± 0.67 | r |
| | | 1031+350 | | | |
| C | 10 31 11.20 | +35 07 39.2 | 3.39 ± 0.08 | 10.16 ± 0.25 | r |
| | | 1037+110 | | | |
| C | 10 37 50.92 | +11 08 58.3 | 4.07 ± 0.08 | 4.54 ± 0.24 | u |
| | | 1049+442 | | | |
| A | 10 49 30.48 | +44 24 43.7 | 6.81 ± 0.07 | 8.61 ± 0.21 | r |
| C | 10 49 30.48 | +44 24 43.3 | 8.76 ± 0.15 | 9.51 ± 0.46 | r |
| | | 1056+244 | | | |
| C | 10 56 36.19 | +24 48 39.4 | 20.70 ± 0.04 | 22.21 ± 0.13 | u |
| | | 1106+271 | | | |
| C | 11 06 56.06 | +27 11 30.9 | 6.48 ± 0.49 | 7.86 ± 0.15 | u |
| | | 1836+354 | | | |
| A | 18 36 49.23 | +35 49 35.9 | 6.82 ± 0.05 | 8.71 ± 0.15 | r |
| C | 18 36 49.22 | +35 49 35.8 | 9.17 ± 0.11 | 9.35 ± 0.35 | u |
| | | 1909+450(a) | | | |
| A | 19 09 38.17 | +45 02 23.8 | 0.22 ± 0.03 | 0.88 ± 0.14 | r |
| C | 19 09 38.16 | +45 02 24.0 | 0.96 ± 0.02 | 1.19 ± 0.04 | u |
| | | 1909+450(b) | | | |
| A | 19 09 40.45 | +45 02 25.5 | 1.04 ± 0.01 | 1.59 ± 0.04 | u |
| C | 19 09 40.47 | +45 02 25.2 | 3.27 ± 0.06 | 5.22 ± 0.20 | r |
| | | 2008−030 | | | |
| A | 20 08 47.06 | −03 08 49.6 | 19.26 ± 0.20 | 23.58 ± 0.63 | r |
| C | 20 08 47.04 | −03 08 49.5 | 24.37 ± 0.07 | 25.35 ± 0.21 | u |
| | | 2041−165 | | | |
| A | 20 41 29.13 | −16 51 10.8 | 10.02 ± 0.05 | 11.19 ± 0.16 | r |
| C | 20 41 29.13 | −16 51 10.9 | 12.18 ± 0.06 | 12.54 ± 0.18 | u |
| | | 2139+362 | | | |
| C | 21 39 36.04 | +36 23 04.9 | 6.13 ± 0.08 | 6.17 ± 0.25 | u |
| | | 2144+000 | | | |
| C | 21 44 17.71 | +00 07 20.4 | 4.33 ± 0.03 | 5.04 ± 0.10 | u |

**Table 2**—*Continued*

| Array | RA (1950) | Dec | $S_{6\,\rm cm}$ Peak mJy | $S_{6\,\rm cm}$ Int. mJy | |
|---|---|---|---|---|---|
| (1) | (2) | (3) | (4) | (5) | (6) |
| | | 2150−062 | | | |
| A | 21 50 27.75 | −06 28 59.4 | 4.17 ± 0.01 | 4.64 ± 0.03 | u |
| C | 21 50 27.75 | −06 28 59.1 | 5.60 ± 0.05 | 6.09 ± 0.14 | u |
| | | 2208+270 | | | |
| C | 22 08 19.98 | +27 03 34.7 | 3.53 ± 0.06 | 5.46 ± 0.17 | u |
| | | 2249−180 | | | |
| A | 22 49 09.08 | −18 08 20.6 | 3.31 ± 0.05 | 3.40 ± 0.15 | r |
| C | 22 49 09.08 | −18 08 20.1 | 3.82 ± 0.04 | 4.41 ± 0.12 | u |
| | | 2332+291 | | | |
| C | 23 32 42.41 | +29 13 25.0 | 3.49 ± 0.05 | 3.71 ± 0.16 | r |
| | | 2336+360 | | | |
| A | 23 36 32.23 | +36 04 31.3 | 7.74 ± 0.08 | 9.82 ± 0.23 | r |
| C | 23 36 32.22 | +36 04 31.4 | 11.58 ± 0.08 | 12.78 ± 0.25 | r |

*Fluxes obtained by summing within boundaries of extended features; see text.

## Table 3. 20 cm Data and Spectral Indices

| | RA (1950) | Dec | $S_{20\,\mathrm{cm}}$ Peak mJy | $S_{20\,\mathrm{cm}}$ Int. mJy | $\log \frac{L_{20\,\mathrm{cm}}}{L_\odot}$ | $q$ | $\alpha$ |
|---|---|---|---|---|---|---|---|
| | (1) | (2) | (3) | (4) | (5) | (6) | (7) |
| | 00 15 23.0 | +54 54 36 | $4.03 \pm 0.20$ | $4.08 \pm 0.64$ | 5.58 | 2.84 | $-0.23 \pm 0.14$ |
| | 00 26 13.0 | +42 51 44 | $17.54 \pm 0.42$ | $22.43 \pm 1.31$ | 6.20 | 2.19 | $-0.51 \pm 0.06$ |
| | 01 10 41.5 | +73 22 23 | $20.64 \pm 0.12$ | $23.27 \pm 0.36$ | 5.53 | 2.54 | $-0.77 \pm 0.01$ |
| | 01 52 09.4 | +52 24 44 | $11.84 \pm 0.12$ | $13.58 \pm 0.37$ | 5.81 | 2.48 | $-0.52 \pm 0.02$ |
| * | 02 48 18.5 | +43 02 53 | — | $20 \pm 5$ | — | — | |
| | 03 15 51.7 | +42 27 34 | $19.59 \pm 0.30$ | $19.39 \pm 0.97$ | 6.42 | 2.44 | $-0.45 \pm 0.04$ |
| | 03 28 49.2 | +01 08 13 | $11.44 \pm 0.24$ | $13.02 \pm 0.76$ | 5.86 | 2.35 | $-0.59 \pm 0.06$ |
| | 03 50 19.1 | −03 28 56 | $10.89 \pm 0.14$ | $12.53 \pm 0.45$ | 5.69 | 2.51 | $-0.47 \pm 0.04$ |
| | 03 52 08.0 | +00 28 17 | $5.02 \pm 0.21$ | $5.96 \pm 0.64$ | 6.01 | 2.78 | $-0.21 \pm 0.09$ |
| | 04 15 28.6 | +17 55 26 | $57.77 \pm 0.71$ | $63.36 \pm 2.12$ | 6.17 | 1.93 | $-0.85 \pm 0.03$ |
| | 04 23 15.2 | +14 36 54 | $28.79 \pm 0.44$ | $31.86 \pm 1.25$ | 6.18 | 2.14 | $-0.80 \pm 0.03$ |
| | 05 24 41.4 | +01 03 31 | $11.38 \pm 0.39$ | $14.42 \pm 1.18$ | 6.00 | 2.30 | $2.13 \pm 0.07$ |
| | 07 38 44.9 | −03 44 40 | $17.73 \pm 0.31$ | $20.67 \pm 0.97$ | 5.67 | 2.35 | $-0.81 \pm 0.06$ |
| | 08 07 08.5 | +05 10 02 | $32.13 \pm 0.39$ | $32.93 \pm 1.18$ | 5.83 | 2.29 | $-0.72 \pm 0.04$ |
| | 08 34 05.3 | +15 50 46 | $7.80 \pm 0.41$ | $8.92 \pm 1.30$ | 5.61 | 2.49 | $-0.73 \pm 0.13$ |
| | 08 57 59.3 | +34 47 15 | $25.91 \pm 0.50$ | $31.55 \pm 1.58$ | 6.01 | 2.11 | $-0.91 \pm 0.06$ |
| | 09 01 48.9 | +14 47 38 | $53.78 \pm 1.74$ | $68.18 \pm 5.27$ | 6.10 | 1.97 | $-0.61 \pm 0.08$ |
| | 09 06 11.4 | −12 48 43 | $16.86 \pm 0.34$ | $22.07 \pm 1.07$ | 5.95 | 2.36 | $-0.77 \pm 0.06$ |
| | 09 11 10.8 | −10 07 04 | $24.08 \pm 0.34$ | $27.88 \pm 1.07$ | 5.80 | 2.55 | $-0.75 \pm 0.05$ |
| | 09 32 04.9 | +61 34 37 | $146.88 \pm 0.47$ | $148.66 \pm 1.43$ | 6.25 | 2.07 | $-0.68 \pm 0.04$ |
| | 09 58 21.5 | +47 14 15 | $21.84 \pm 0.72$ | $24.42 \pm 2.19$ | 6.13 | 2.14 | $-0.41 \pm 0.07$ |
| | 10 03 36.0 | +48 52 28 | $19.83 \pm 0.35$ | $28.51 \pm 1.09$ | 5.95 | 2.34 | $-0.65 \pm 0.05$ |
| | 10 19 02.3 | +13 22 05 | $16.46 \pm 0.45$ | $19.71 \pm 1.31$ | 5.94 | 2.38 | $-0.89 \pm 0.10$ |
| | 10 31 11.2 | +35 07 41 | $27.47 \pm 0.26$ | $33.36 \pm 0.81$ | 6.10 | 2.08 | $-0.99 \pm 0.03$ |
| | 10 37 50.9 | +11 08 59 | $7.64 \pm 0.37$ | $7.89 \pm 1.12$ | 6.04 | 2.51 | $-0.46 \pm 0.13$ |
| | 10 49 30.5 | +44 24 45 | $22.59 \pm 0.28$ | $23.53 \pm 0.84$ | 6.17 | 2.31 | $-0.75 \pm 0.05$ |
| | 10 56 36.2 | +24 48 39 | $46.40 \pm 0.32$ | $45.78 \pm 0.92$ | 5.80 | 2.53 | $-0.60 \pm 0.02$ |
| | 11 06 56.0 | +27 11 32 | $18.04 \pm 0.13$ | $19.57 \pm 0.38$ | 5.86 | 2.22 | $-0.76 \pm 0.02$ |
| | 18 36 49.2 | +35 49 36 | $20.91 \pm 0.34$ | $24.02 \pm 0.11$ | 6.38 | 2.13 | $-0.78 \pm 0.03$ |
| (a) | 19 09 38.1 | +45 02 25 | $2.62 \pm 0.14$ | $4.66 \pm 0.46$ | 5.15 | 2.93 | $-1.13 \pm 0.09$ |
| (b) | 19 09 40.4 | +45 02 25 | $12.26 \pm 0.24$ | $15.10 \pm 0.77$ | — | — | $-0.88 \pm 0.05$ |
| | 20 08 47.1 | −03 08 49 | $53.59 \pm 0.35$ | $52.34 \pm 1.05$ | 6.64 | 2.07 | $-0.60 \pm 0.05$ |
| | 20 41 29.1 | −16 51 10 | $19.84 \pm 0.16$ | $21.05 \pm 0.49$ | 6.08 | 2.44 | $-0.43 \pm 0.02$ |
| | 21 39 36.0 | +36 23 05 | $14.73 \pm 0.27$ | $16.23 \pm 0.86$ | 6.06 | 2.27 | $-0.80 \pm 0.06$ |
| | 21 44 17.7 | +00 07 20 | $11.23 \pm 0.22$ | $12.64 \pm 0.71$ | 5.71 | 2.39 | $-0.76 \pm 0.05$ |
| | 21 50 27.6 | −06 29 00 | $9.30 \pm 0.60$ | $15.31 \pm 1.82$ | 5.84 | 2.44 | $-0.77 \pm 0.10$ |
| | 22 08 20.0 | +27 03 34 | $13.55 \pm 0.06$ | $14.64 \pm 0.18$ | 5.86 | 2.27 | $-0.82 \pm 0.03$ |
| | 22 49 09.0 | −18 08 20 | $6.71 \pm 0.12$ | $7.13 \pm 0.38$ | 5.51 | 2.95 | $-0.40 \pm 0.05$ |
| | 23 32 42.4 | +29 13 25 | $7.83 \pm 0.24$ | $7.30 \pm 0.68$ | 5.79 | 2.56 | $-0.56 \pm 0.09$ |
| | 23 36 32.2 | +36 04 31 | $24.06 \pm 0.22$ | $23.88 \pm 0.66$ | 5.87 | 2.60 | $-0.52 \pm 0.03$ |

* The position of 0248+43 was inferred from the 6-cm maps.

## Table 4: Radio Flux Density Measurements from the Literature

| Source | Ref. | RA | Dec | *$S_{20}$ | $S_6$ | $S_{60}$ | $S_{100}$ | $cz$ | $\log \frac{L_{FIR}}{L_\odot}$ |
|---|---|---|---|---|---|---|---|---|---|
| | | (1950) | (1950) | mJy | mJy | Jy | Jy | km s$^{-1}$ | |
| (1) | (2) | (3) | (4) | (5) | (6) | (7) | (8) | (9) | (10) |
| IRAS 0033−273 | U | 00 33 31.9 | −27 32 04 | 13.9 | | 4.42 | 3.17 | 20771 | 11.64 |
| IRAS 0045−290 | U | 00 45 40.6 | −29 04 38 | 18.7 | | 2.55 | 3.48 | 33060 | 11.88 |
| IRAS 0100−224 | U | 01 00 22.8 | −22 38 09 | 5.2 | | 2.24 | 1.79 | 35286 | 11.81 |
| IRAS 0126+330 | S | 01 26 24.1 | +33 02 52 | | 5 | 2.20 | 2.82 | 24145 | 11.53 |
| IRAS 0136−104 | C | 01 36 24.3 | −10 42 25 | 15.4 | | 6.74 | 6.59 | 14250 | 11.52 |
| Mrk 1014 | A | 01 57 16.6 | +00 09 07 | 22.9 | 7.92 | 2.34 | 2.29 | 48869 | 12.13 |
| | N | | | | 8.2 | | | | |
| IRAS 0518−252 | C | 05 18 55.9 | −25 24 39 | 28.1 | | 13.18 | 11.99 | 12760 | 11.71 |
| IRAS 0857+391 | C | 08 57 13.0 | +39 15 39 | 6.6 | | 7.23 | 4.96 | 17480 | 11.69 |
| | N | | | | 5.3 | | | | |
| IRAS 1017+082 | C | 10 17 22.3 | +08 28 40 | 9.4 | | 5.96 | 5.67 | 14390 | 11.48 |
| NGC 3690 | C | 11 25 43 | +58 50 16 | 667 | 300 | 119.7 | 118.6 | 3101 | 11.45 |
| | N | | | | 200 | | | | |
| IRAS 1211+030 | C | 12 11 12.37 | +03 05 19.7 | 23.8[a] | | 8.36 | 9.91 | 21703 | 12.01 |
| | C | 12.48 | 22.1 | | | | | | |
| | N | | | | 13 | | | | |
| 3C 273 | K | 12 26 33.2 | +02 19 43 | 45000 | 45000 | 2.22 | 2.91 | 47469 | 12.13 |
| Mrk 231 | C | 12 54 05.0 | +57 08 38 | 296 | | 33.57 | 30.9 | 12518 | 12.10 |
| | N | | | | 100 | | | | |
| Mrk 273 | C | 13 42 51.71 | +56 08 14.3 | 145 | 103 | 23.71 | 22.28 | 11180 | 11.85 |
| | N | | | | 70 | | | | |
| IRAS 1345+123 | NED | 13 45 06.2 | +12 32 20 | 5215 | 3060 | 2.01 | 2.14 | 36341 | 11.82 |
| IRAS 1345−300 | vD | 13 45 28.4 | −29 56 58 | | 5.2 | 2.34 | 3.81 | 22728 | 11.54 |
| IRAS 1434−144 | C | 14 34 53.20 | −14 47 26.1 | 38.1[a] | | 6.82 | 7.50 | 24677 | 12.02 |
| | C | 53.42 | 23.1 | | | | | | |
| IZw 107 | C | 15 16 19.30 | +42 55 38.3 | 48.5[a] | | 9.38 | 10.26 | 12049 | 11.54 |
| | C | 19.51 | 31.3 | | | | | | |
| IRAS 1525+360 | C | 15 25 03.7 | +36 09 01 | 13.8 | | 7.50 | 5.86 | 16602 | 11.68 |
| | N | | | | 11.8 | | | | |
| Arp 220 | C | 15 32 46.88 | +23 40 07.9 | 326.6[a] | 204 | 107.4 | 118.3 | 5534 | 11.92 |
| | C | 46.95 | 07.7 | | | | | | |
| | S91 | | | | 210 | | | | |
| NGC 6240 | C82 | 16 50 27.8 | +02 28 58 | | 80 | 22.54 | 27.29 | 7298 | 11.49 |
| | CWB | | | 390 | | | | | |
| IRAS 1713+531 | C | 17 13 13.6 | +53 13 50 | 26.2 | | 6.30 | 7.98 | 15270 | 11.59 |
| IRAS 1720−001 | B | 17 20 48.2 | −00 14 17 | | 69 | 34.12 | 35.73 | 12836 | 12.15 |
| IRAS 2320+060 | S | 23 20 28.4 | +06 01 38 | | 7.5 | 4.27 | 4.25 | 16779 | 11.47 |

* corrected to our frequency of 1460 MHz using the spectral indices of Condon et al. (1991) or other estimates of the spectral index, when available. For other sources $\alpha = -0.7$ is used.

[a] listed flux includes that of the next component listed.

References:

A = Antonucci & Barvainis 1988 (20 and 6 cm, C array);
B = Becker et al. 1991 (6 cm);
C = Condon et al. 1990 (20 cm, B and C arrays), Condon et al. 1991c (20 and 6 cm, A array) and Condon et al. 1991b (6cm, D array);
C82 = Condon et al. 1982 (20 and 6 cm, A array);
CWB = Colbert, Wilson, & Bland-Hawthorn 1994 (20 cm, B array);
K = Kellerman et al. 1969 (single dish)
N = Neff & Hutchings 1992 (6 cm, A array);
NED = NASA/IPAC Extragalactic Database (based on single dish);
S = Sopp & Alexander 1992 (6 cm, D array);
S91 = Sopp & Alexander 1991b (6 cm, 3 arrays);
U = Unger et al. 1989 (20 cm, hybrid arrays);
vD = van Driel, van den Broek, & de Jong 1991 (6 cm, C/D array).